\documentclass[pre,preprint,aps,]{revtex4}
\usepackage{graphicx}
\usepackage{comment}
\usepackage{amsmath,amssymb,amsthm} 
\usepackage{bm}
\usepackage{verbatim}
\usepackage{float}

\begin{document}

		\title{ The role of the attractive forces in a  supercooled liquid}

\author{ S\o ren  Toxvaerd }
\affiliation{DNRF centre  ``Glass and Time,'' IMFUFA, Department
 of Sciences, Roskilde University, Postbox 260, DK-4000 Roskilde, Denmark}
\date{\today}

\vspace*{0.7cm}

\begin{abstract}
 Molecular Dynamics simulations of crystallization in a supercooled liquid of Lennard-Jones particles with
different range of attractions shows that the inclusion of the attractive forces from  the first, second and third coordination shell
 increases the trend to crystallize systematic. The bond-order $Q_6$ in the supercooled liquid
 is heterogeneously distributed with  clusters of particles with relative high bond-order for a supercooled liquid, and a systematic
increase  of the extent of  heterogeneity with increasing range of attractions. The onset of
crystallization appears in such a cluster, which  together explains the attractive forces influence on  crystallization.
The mean square displacement and self-diffusion constant exhibit the same dependence on the range of attractions
in the dynamics and shows, that the attractive forces and the range of the forces plays an important role
for bond-ordering, diffusion and for crystallization.
\end{abstract}
\maketitle

\section{Introduction}
 Ever since Berni Alder \cite{Alder} in 1957  performed the first Molecular Dynamics (MD) simulation of a hard sphere system with
 crystallization there has been a general understanding of, that crystallization of a simple liquid is given
 by the harsh repulsive short range forces, and the MD simulation with particles with the more realistic Lennard-Jones (LJ) potential 
 supported the assumption \cite{Verlet,Hansen}. Not only did the thermodynamics  of a LJ system agreed with the corresponding behaviour of a system of noble  gas atoms,
 but the tendency to crystallize for LJ systems  with  and without  attractive forces is also very similar (Figure 1). This similarity is explained with,
 that there is an overall agreement between the radial distribution function $g(r)$ for LJ systems with-
 and without the attractive forces, as well as with the radial distribution function for a hard-sphere system.
These  rather closely similarities  in the radial distributions of the particles  have given reason to the well-established ``perturbation theory" \cite{Barker,WCA}, where the thermodynamic and dynamic behavior of
 a system is obtained from systems of purely repulsive particles  by mean field corrections for  contributions from the attractive forces.

Here we analyse the role of the attractive forces on the supercooled state and the crystallization  by Molecular
Dynamics (MD) simulations of  LJ systems with different range of attractions. The simulations
show, that the attractive forces play an important role in a supercooled liquid. They increase the bond-order in the supercooled liquid, given by $Q_6$ \cite{Steinhardt,Dellago}, and the
tendency to crystallize.

\begin{figure}
	\begin{center}
		\includegraphics[width=5cm,angle=-90]{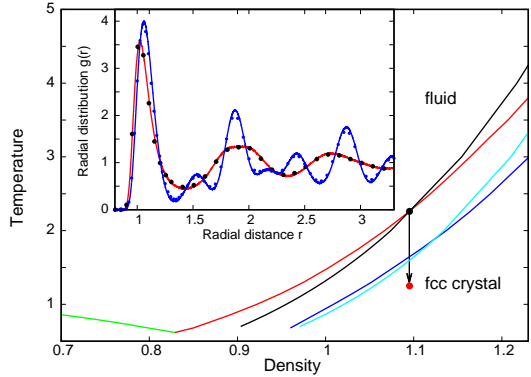} 
	\end{center}
\caption{The liquid-solid phase diagram for a LJ and a WCA system. LJ: green is liquid in equilibrium with gas;
red: liquid in equilibrium with fcc solid: blue. WCA: black is fluid in equilibrium  with fcc solid: light blue.
	The systems are cooled down from  liquids at  $(\rho, T$) =(1.095, 2.25): black point   to  $(\rho, T$) =(1.095, 1.25): red point.
Inset: the radial distribution functions g(r) at (1.095, 1.25).  Red: supercooled   liquid,
and with black points for WCA. Blue: fcc crystal and with blue points WCA.}
\end{figure}\

\section{Crystallization of the supercooled liquids}

The systems with different range of attractions are cooled down from the liquids at the state point $(\rho_l, T_c)$ =(1.095, 2.25) for a liquid
in equilibrium with fcc solid.
The $(\rho, T$) phase diagrams for LJ systems with- and without attractive forces
are shown in Figure 1. The  $T(\rho)$ curves for coexisting liquids and fcc solids are obtained 
thermodynamically \cite{Sadus,Barroso}. At  $(\rho_l, T_c)$ =(1.095, 2.25) a liquid with only repulsive forces and the LJ
liquid with (full) attractions crystallize at the same state point.

 The MD systems consist  of $N=80000$ LJ  particles,
 where the forces are ``cutted and shifted"
 to zero at different  particle distances greater than $r_c$ \cite{tox1}.
The  simulations (unit length $l^*: \sigma$;  unit time $t^*: \sigma \sqrt{m/\epsilon}$, for computational details
see  \cite{tox1}) are performed for four different values of $r_c$: 3.5, 2.3, 1.41
and $2^{1/6}$, respectively.
Only the  strong repulsive LJ forces are included in
the dynamics for the short cut at $r_c=2^{1/6}$, and this system appears
in the literature with the name WCA  \cite{WCA}.  For $r_c=1.41$ also the attractive forces from particles in the
first coordination shell
are included,
for $r_c=2.3$  the forces from the second coordination shell are included, and for $r_c=3.5$ the
third shells forces are
include in the dynamics.


There is a remarkably similarity between the LJ- and the WCA system,
which had led to to the perturbation expansion theories  
\cite{Barker,WCA}. This is caused by  the similarity in the radial distributions, and the inset shows the radial
distribution function $g(r)$ for the two systems in the supercooled state
(red point in the Figure). With red is $g(r)$ for the supercooled liquid  and the black points are $g(r)$ for the WCA system.
The blue curve in the inset is $g(r)$ for a LJ fcc crystal  and with blue points for fcc WCA.
   The  $g(r)$ for the WCA systems  are shown with points for illustrative reasons because the differences between  $g(r)$
    for LJ and WCA are small.
The overall  similarity between the two  systems $g(r)$ implies, that the 
mean pair-distributions in the supercooled state with coordination shells around a particles are almost identical for the two systems,
and that the mean effects of the attractive forces on pressure, energy and free energy can be obtained as 
mean field contributions.

 The  latent heat is released when a supercooled liquid crystallizes spontaneously, and the energy decreases and
  the temperature  increases without a thermostat. In \cite{tox2} the effect of a thermostat on the spontaneous crystallizations in the big MD 
   supercooled systems
    was investigated by performing ensemble simulations with- and without a thermostat and with the conclusion, that the intensive MD
     thermostat, as expected,  had no effect on the onset of crystallization. The present ($NVT$) simulations are with a thermostat  by which
      the latent heat is  removed smoothly, and the energy per particle  decreases during the spontaneous
      crystallization at the constant  supercooled temperature (Figure 2).

The systems are cooled down from  liquids at the state point 
 where the  liquids are in thermodynamic equilibrium with fcc solid,
 and at the point on the coexisting phase lines where the   lines for the two
systems crosses each other, by which the
degree of supercooling  $T/T_c=1.25/2.25=0.556$  at the constant density ($\rho_l=1.095$) is the same for the systems.
The systems crystallize, however, with different tendency as can be seen in Figure 2.

\begin{figure}
	\begin{center}
		\includegraphics[width=5cm,angle=-90]{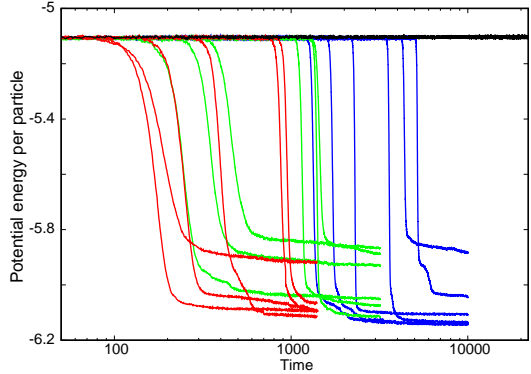} 
	\end{center}
		\caption{Energy per particle as a function of $log-$time after the quench to the supercooled state.
		With black lines are for the systems with only repulsive forces (WCA). The six blue curves are for
		$r_c$=1.41 (with  the attractive forces in the  first coordination shell included);
		green six curves are for $r_c$=2.3 (with attractive forces also from second coordination shell);
		red six curves $r_c$=3.50 (with  attractive forces also from third coordination shell).}		
\end{figure}

\begin{figure}
	\begin{center}
		\includegraphics[width=5cm,angle=-90]{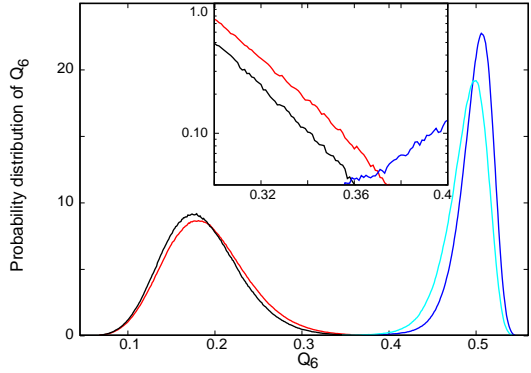} 
	\end{center}
		\caption{Probability distribution P($Q_6$) of  bond-order  $Q_6$ in the supercooled state. Red: LJ supercooled liquid with $r_c=3.5$; black: supercooled WCA fluid;
			blue: fcc  LJ with $r_c=3.5$; light blue: WCA fcc. Inset: $log$(P)  in the interval $Q_6 \in [0.30, 0.40]$.}
\end{figure}

Figure 2  shows the $log-$time evolution  of the energies per particle  for systems in the supercooled state and  for different 
values of $r_c$. The  mean field energies from particles in the interval $[r_c, 3.5]$ are added to the functions
\begin{equation}
	u(t)= u(t,r_c)+ 2 \pi \rho \int_{r_c}^{3.5}   g(r)u_{\textrm{LJ}}(r)r^2 dr,
\end{equation}
and the energies $u(t)$ for  different cuts of the forces are almost equal before the onset of crystallization in accordance  with the perturbation theory.
The time evolution are shown with a logarithmic time scale.
With black lines are the WCA systems with only repulsive forces, and they were simulated $ \Delta t$= 22000 (2.2 $\times 10^{7}$ time steps).
However, they remained  in the
supercooled state without crystallization. (The WCA systems were crystallized at a lower temperature $T$=1.15.) The six blue curves are for
$r_c$=1.41  with  the attractive forces in the  first coordination shell included in the force, and they crystallized within the time interval $ [1200, 5000]$ after
the supercooling. The
  green curves are for $r_c$=2.3 with attractive forces also from the second coordination shell, and they crystallized  within the time interval $ [150, 1300]$.  
The   red curves are for  $r_c$=3.50 with  attractive forces also from  the third coordination shell, and they  crystallized  within the time interval $ [80, 850]$.
These data indicate a logarithmic effect of the attractive forces on the stability of a supercooled liquid. The effect of the attractive forces
for $3.5 < r_c$ on the stability of the supercooled state was, however, not investigated due to a lack of computer facilities for these very demanding simulations.  

(The energies after the crystallization are rather different. In general a hard sphere system as well as a LJ system crystallizes
with polymorphism \cite{Snook,Delhommelle1,Delhommelle2} to different polycrystalline fcc states with different
mean energy per particles \cite{tox2}, as also seen in Figure 2.)

 The sensitivity of the range of attractions  to the ability to crystallize is surprising given that  the pair distributions of the different systems are
 very similar in the
 supercooled state as well as in the crystalline state. In the classical nucleation theory  the size
 of the critical nucleus  is the size, where the gain in free energy by an increase of particles in
 the crystal nucleus equals the cost of the increasing surface free energy, and these excess
 free energies should not be sensitive to the range of attractions due to the
 similarities in $g(r)$. But the distribution of  bond-order  $Q_6$ for the particles is sensitive to
 the range of the attractive forces.

\begin{figure}
	\begin{center}
		\includegraphics[width=5cm,angle=-90]{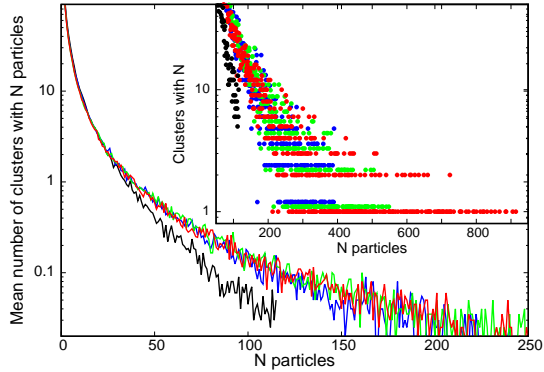} 
	\end{center}
	\caption{Distributions (logarithmic) of clusters of $N$ particles with $Q_6 >0.25$ in the supercooled LJ liquids. 
	Red: for the LJ system with $r_c$=3.5; green: $r_c$=2.3; blue: $r_c$=1.41; black: $r_c=2^{1/6}$ (WCA).
	Inset: The discrete distributions of big clusters (with points). }
\end{figure}

A supercooled  LJ liquid is characterized by a heterogeneous distribution  of bond-order, given by $Q_6$ \cite{tox2}. Here we will argue,
that  it is the attractive forces impact on the extent of  the heterogeneity of the bond-order, which causes the difference in the tendency to crystallize.
 The   distribution  $P(Q_6)$ of bond order $Q_6(i)$ for the particles $i$  in the LJ supercooled state is shown in Figure 3 and with an inset,
 which shows the $log-$distributions in the bond-order interval for which the  particles with these bond-order are  heterogeneous distributed. As in \cite{tox2}  there is an overlap in the distributions
 in supercooled liquid and in fcc crystal in this $P(Q_6)$ interval $Q_6 \in [0.35, 0.38]$.  
In \cite{tox2} we found, that the particles with a relative high liquid bond-order $Q_6 > 0.25$ were heterogeneous distributed and with some particles 
 with bond-order  $Q_6 > 0.35$.  Furthermore the critical crystal nucleus appeared in such a domain and with mean bond-order  $<Q_6> \approx 0.38$,
which is significantly less than the bond-order in the fcc crystal at the same state point.

The number of clusters with $N$ particles with a bond-order $0.25 < Q_6 $ are shown in the next figure. (A particle $i$ in a cluster with $Q_6(i) > 0.25$ is close to 
($r_{ij} < 1.41$)  at least one other particle $j$ in the cluster.) The distributions are obtained for the supercooled state as the mean of 200
independent determinations, and the inset shows  the number of discrete and rare events of bigger clusters. The figure shows two things. For the first there is a crucial difference between
the purely repulsive force system (WCA)  and the systems with attractive forces which all exhibit big clusters with high liquid bond-order.
And secondly, the inset
shows, that although the three distributions with different range of attractions looks pretty similar,
there  appears occasionally a  much bigger clusters for
the systems with long range attractions.

The critical nucleus  were determined as described in \cite{tox2}. Figure 5 shows a representative example of the time evolution of
the number of particles $N(t)$ in the biggest cluster,
and with the mean bond-order in the inset of the figure. The estimated critical  sizes  $<N_c>$ with the bond-order $<Q_6>$ for the simulations are:\\
$r_c=1.41: <N_c>= 73 \pm 3$ and $<Q_6>=0. 390 \pm 0.007$\\
$r_c=2.30:  <N_c>= 73 \pm 5 $ and $<Q_6>=0. 389 \pm 0.001$\\
$r_c=3.50: <N_c>= 73  \pm 6$ and $<Q_6>=0. 392 \pm 0.007 $.\\

\begin{figure}	
\begin{center}
\includegraphics[width=5cm,angle=-90]{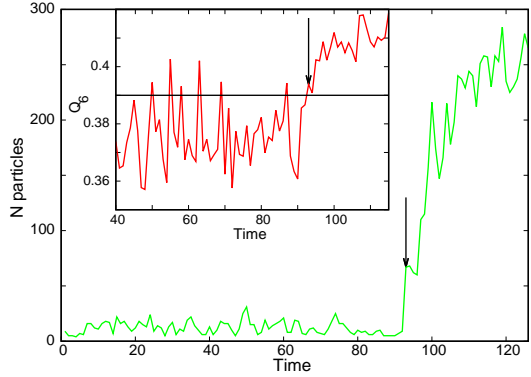} 
\end{center}
	\caption{An example of the number $N(t)$ of particles in the biggest cluster of particles $i$ with $Q_6(i) >0.25$. The figure shows $N(t)$ at the onset of
	crystallization for a system with $r_c=2.3$ (the green curve to the right in Figure 2).
The inset shows the mean bond-order in the cluster. The arrows point to the time where the critical nucleus appears with $N_c$=0.67 and $<Q_6>$=0.3938.}
\end{figure}

The mean-bond order $<Q_6>=0.390$ in the critical nucleus is much less than in an ordered fcc crystal and
the size of critical nuclei   is the same for the three ranges of attractions. The extension of domains with relative  high
bond-order  varies, however, with the range of the cut (Figure 4), and the critical nuclei appear in a domain with high bond-order. So 
the extend of the heterogeneous distribution of high bond-order in the
supercooled liquid  and thereby the probability to obtain a critical nucleus
 can explain the observed differences in the tendency to crystallize.

The existence of ``dynamic heterogeneity" in supercooled liquids have been known for a long time \cite{Ediger,Harrowell1,Harrowell2}, and in 
\cite{tox2} is was  linked to the bond-ordering in subdomains. If so the viscosity and particle diffusion should be different for domains with relative low bond-order
compared with subdomains with relative high bond-order. This is, however, difficult to determine directly because the domains are not permanent and particles change
bond-order with time. But the overall particle diffusion reveal the differences and the main effect of the attractive forces on diffusion and viscosity.
The next figure gives the mean square displacements of
a particle in the supercooled state and for different range of attractions. The figure shows that there is a difference in the slopes end thereby the self-diffusion constants $D$
for the different range of attractions. The self-diffusion constants are WCA: $D$=0.01315;  $r_c =1.41$: $D$=0.01033;  $r_c=2.30$: $D$=0.00991;  $r_c=3.5$: $D$=0.01004.
The inset is the mean square displacements in logarithmic scales, and it shows that the short
time ``ballistic regime" is similar  for all four systems and given by the strong repulsive forces.
The behaviour of the particle diffusion with respect to the range of the attractions  can be explained by,
that the  domains with relative high bond-order slow down the
 mobility of a particle in these domains, i.e.  that the   mobility is heigh  in domains where the bond-order is small and small  in domains with relative high bond-order, where the particles 
  are  tied together weakly. 

The attractive forces dynamic effect
in  supercooled states has also been obtained for mixtures \cite{ber11,toxdyre}

\begin{figure}	
\begin{center}
\includegraphics[width=5cm,angle=-90]{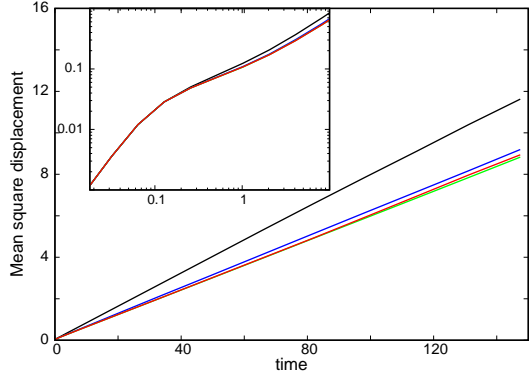} 
\end{center}
\caption{Mean square displacements at ($T, \rho)$ = (1.25, 1.095). Black: for $r_c=2^{1/6}$(WCA); blue: $r_c$=1.41;  green: $r_c$=2.3; and red:  $r_c$=3.5, respectively. Inset: In logarithmic scales.}
\end{figure}

\section{Conclusion}
$In $ $conclusion$ 
 The radial distribution functions $g(r)$ for the supercooled liquid as well as  for the fcc solid
are insensitive to  the range of  the attractions, and hence the
free energies per particle (chemical potential) are also insensitive. Consistent with this observation, so is the size of the critical nuclei;
but nonetheless, the tendency to crystallize
depends on the range of attraction.
The systems in the supercooled state exhibit, however, heterogeneous distributed particles with  a relative high bond-order for the supercooled state, and
the extent of the heterogeneity is enhanced  mainly from the attractions from the particles within the first coordination shell, but also  the particles
from the second- and third coordination shells increase the number of  domains with relative high bond-order. In accordance with this observation, the
systems crystallize much more easily for the systems with attractions and in a systematic way so that six crystallizations  of
particles with attractions   from particles  within  three coordination shells
crystallized $\approx $ eight times faster than six systems with  attractions only from the first  coordination shell,
whereas the  systems with only repulsive forces did not crystallized at the supercooled state point.

This behavior  of the heterogeneous bond-order distribution is consistent with the well known ``dynamic heterogeneity" in supercooled liquids, and the  self-diffusion  for the particles
with different range of attractions supports the hypothesis. The attractions slow down the self-diffusion, the main effect comes from the attractions within the first coordination shell,
but also the longer-range  attractions affect the diffusion.  So in summary the attractive forces enhance the extent of the domains with high  bond-order, slow down the particle diffusion and
catalyze the crystallization. The sensitivity of the  crystallization to the range of attractions makes it difficult to compare  nucleation rates obtained by simulations with experimentally determined nucleation rates.

This work was supported by the VILLUM Foundation's Matter project grant No. 16515.


\begin{thebibliography}{99}
	\bibitem{Alder} B. J. Alder and T. E. Wainwright, J. Chem. Phys.  {\bf 27}, 1208 (1957).
	\bibitem{Verlet} L. Verlet Phys. Rev. {\bf 159}, 98 (1967).
	\bibitem{Hansen} J. P Hansen and L. Verlet, Phys. Rev. {\bf 184}, 151 (1969).
	\bibitem{Barker} J. A. Barker and D. Henderson, J. Chem. Phys.  {\bf 47}, 4714 (1967).
	\bibitem{WCA} J. Weeks, D. Chandler and H. C. Andersen, J. Chem. Phys.  {\bf 54}, 5237 (1971).
	\bibitem{Steinhardt} P. J. Steinhardt, D. R. Nelson and M. Ronchetti,  Phys. Rev. B,  {\bf 28}, 784 (1983).
	\bibitem{Dellago} W. Lechner and C. Dellago, J. Chem. Phys.  {\bf 129}, 114707 (2008).
	\bibitem{Sadus} A. Ahmed and R. J. Sadus, Phys. Rev E  {\bf80}, 061101 (2009).
	\bibitem{Barroso} M. A. Barroso and A. L. Ferreira,  J. Chem. Phys. {\bf 116}, 7145 (2002).
	\bibitem{tox1}  S. Toxvaerd and  J. C. Dyre, J. Chem. Phys. {\bf 134}, 081102 (2011).		
%
	\bibitem{tox2} S. Toxvaerd, Eur. Phys. J. B {\bf 93}, 202 (2020).		
	\bibitem{Ediger} M. D. Ediger, Annu. Rev. Phys. Chem. {\bf 51}, 99 (2000).
	\bibitem{Harrowell1} A. Widmer-Cooper, P. Harrowell and H.   Fynewever,	
			Phys. Rev. Lett. \textbf{93}, 135701 (2004).
	\bibitem{Harrowell2} A. Widmer-Cooper and P. Harrowell,
			Phys. Rev. Lett. \textbf{96}, 185701 (2006).
	\bibitem{Snook} B. O'Malley and I.  Snook, 	
				Phys. Rev. Lett. \textbf{90}, 085702 (2003).
	\bibitem{Delhommelle1} C. Desgranges and J. Delhommelle, Phys. Rev. Lett. \textbf{98}, 235502 (2007).
	\bibitem{Delhommelle2} C. Desgranges and J. Delhommelle, J. Phys. Chem. B \textbf{111}, 1465 (2007).		
	\bibitem{ber11} L. Berthier and G. Tarjus, J. Chem. Phys.  {\bf 134}, 214503 (2011).
	\bibitem{toxdyre} S. Toxvaerd and J. C. Dyre,  J. Chem. Phys.  {\bf 135}, 134501  (2011).

\end{thebibliography}
\end{document}